\documentclass[reprint,amsmath,amssymb,aip,superscriptaddress,nofootinbib]{revtex4-1}

\usepackage{graphicx}
\usepackage{dcolumn}
\usepackage{mathtools}
\usepackage{color}
\usepackage[symbol]{footmisc}

\begin{document}

\title{Accurate closed-form solution of the SIR epidemic model}
\author{Nathaniel S. Barlow}
\affiliation{School of Mathematical Sciences, Rochester Institute of Technology, Rochester, NY 14623, USA}
\author{Steven J. Weinstein}
\affiliation{School of Mathematical Sciences, Rochester Institute of Technology, Rochester, NY 14623, USA}
\affiliation{Department of Chemical Engineering, Rochester Institute of Technology, Rochester, NY 14623, USA}

\date{\today}

\begin{abstract}
An accurate closed-form solution is obtained to the SIR Epidemic Model through the use of Asymptotic Approximants (Barlow et. al, 2017, Q. Jl Mech. Appl. Math, 70 (1), 21-48).  The solution is created by analytically continuing the divergent power series solution such that it matches the long-time asymptotic behavior of the epidemic model.  The utility of the analytical form is demonstrated through its application to the COVID-19 pandemic. 

\end{abstract}

\maketitle




There are several problems of mathematical physics in which the only available analytic solution is a divergent and/or truncated power series expansion.  Over the past decade, a new approach has evolved to overcome the convergence barrier in series solutions.   An asymptotic approximant is a closed-form expression whose expansion in one region is exact up to a specified order and whose asymptotic equivalence in another region is enforced.  The remarkable feature of asymptotic approximants is their ability to attain uniform accuracy not only in these two regions, but also at all points in-between, as demonstrated thus far for problems in thermodynamics, astrophysics, and fluid dynamics~\cite{BarlowJCP,BarlowAIChE,Barlow2015,Barlow:2017,Barlow:2017b,Beachley,Belden}.  The current need to model and predict viral epidemics motivates us to extend the application of asymptotic approximants to the commonly used Susceptible-Infected-Recovered (SIR) model.  This model is formulated as a system of nonlinear ordinary differential equations.  Although no exact analytic solution has yet been found for the SIR model, a convergent series solution may be formulated via the homotopy analysis method~\cite{Khan}.  Here, we provide an alternative and simple analytic approach.  Interestingly, the SIR model shares the same asymptotic features as boundary layer flow over a moving flat plate, for which asymptotic approximants have already been applied~\cite{Barlow:2017}.  The analytic nature of the asymptotic approximant derived in what follows is advantageous.   Model parameters may be extracted for available COVID-19 data via a least squares (or equivalent) technique without the need for an embedded numerical scheme.

The SIR epidemic model considers the time-evolution of a susceptible population, $S(t)$, interacting with an infected population, $I(t)$, where $t$ is time.  This model is expressed as~\cite{Kermack}
\begin{subequations}
\begin{equation}
\frac{dS}{dt}=-rSI
\label{eq:S}
\end{equation}
\begin{equation}
\frac{dI}{dt}=rSI-\alpha I
\label{eq:I}
\end{equation}
with constraints
\begin{equation}
S=S_0,~I=I_0\text{ at }t=0,
\label{eq:constraints}
\end{equation}
\label{eq:SIR}
\end{subequations}
where $r$, $\alpha$, $S_0$, $I_0$ are non-negative constant parameters~\cite{Kermack}.  Once~(\ref{eq:SIR})  is solved, the recovered population is extracted as:
\begin{equation}
R(t)=\alpha\int_0^t I(\zeta) d\zeta.
\label{eq:R}
\end{equation}
Equation~(\ref{eq:S}) can be thought of as a standard collision model in a 2${}^\mathrm{nd}$-order chemical reaction, where species $S$ and $I$ ``collide'' to deplete the population of $S$ to create the species $I$.  In this interpretation, \textit{r} is a rate constant, which in practice may be reduced by population behavior such as ``social distancing''.  In the case where $\alpha$=0 in~(\ref{eq:I}), the system~(\ref{eq:SIR}) indicates that $S+I=S_{0} +I_{0} $ for all time.  For \textit{$\alpha \ne 0$}, then, the number of infected are reduced in time in accordance with~(\ref{eq:I}), and it is seen that the parameter \textit{$\alpha$} determines the rate of recovery of infected individuals.  The omission of a negative $\alpha I(t)$ term in~(\ref{eq:S}) is an implicit model assumption that the recovered population is no longer susceptible to the disease.  

We now manipulate the system~(\ref{eq:SIR})  into an equivalent first-order equation to simplify the analysis that follows.  Equations~(\ref{eq:S}) and~(\ref{eq:I}) are divided to obtain
\begin{equation} 
\frac{dI}{dS} =\frac{\alpha }{rS} -1.
\label{eq:divide} 
\end{equation} 
Subsequent integration of~(\ref{eq:divide}) with respect to $S$ and application of the constraints~(\ref{eq:constraints}) yields
\begin{equation}
I=\frac{\alpha }{r} \ln \left(\frac{S}{S_{0} } \right)-S+S_{0} +I_{0} .  
\label{eq:algebraic}
\end{equation} 
Equation~(\ref{eq:algebraic}) is substituted into equation~(\ref{eq:S}) to obtain
\begin{subequations}
\begin{equation}
\frac{dS}{dt} =\beta S+rS^{2}-\alpha S\ln S 
\label{eq:singleODE}
\end{equation}
where
\begin{equation}
 \beta =\alpha \ln S_{0} -r(S_{0} +I_{0} ).
 \label{eq:beta}
 \end{equation}
From equation~(\ref{eq:constraints}), the constraint on $S$ is:
\begin{equation}
S=S_0\text{ at }t=0.
\label{eq:Sconstraint}
\end{equation}
\label{eq:newODE}
\end{subequations}
System~(\ref{eq:newODE}) is equivalent to~(\ref{eq:SIR}) to solve for $S$ and, once solved, the solution for $I$ may be obtained using~(\ref{eq:algebraic}), which may be integrated to find $R$ from~(\ref{eq:R}).

The series solution of~(\ref{eq:newODE}) is given by\footnote[1]{The original manuscript omitted the $(j+1)$ factor in $b_n$.  We thank C. Reinberger for bringing this to our attention}
\begin{subequations}
\begin{equation}
S=\sum_{n=0}^\infty a_n t^n,~a_0=S_0
\label{eq:series}
\end{equation}
\begin{equation}
a_{n+1}=\frac{1}{n+1}\left[\beta a_n+\displaystyle\sum_{j=0}^na_j\left(ra_{n-j}-\alpha b_{n-j}\right)\right],
\label{eq:coefficients}
\end{equation}
\begin{equation}
b_{n>0}=\frac{1}{n}\sum_{j=0}^{n-1}(j+1)a_{j+1}\tilde{a}_{n-1-j},~~b_0=\ln a_0,
\label{eq:b}
\end{equation}
\begin{equation}
\tilde{a}_{n>0}=\frac{-1}{a_0}\sum_{j=1}^na_j\tilde{a}_{n-j},~~\tilde{a}_0=\frac{1}{a_0}.
\label{eq:atilde}
\end{equation}
\label{eq:SeriesSolution}
\end{subequations}
The result~(\ref{eq:SeriesSolution}) is obtained by applying Cauchy's product rule~\cite{Churchill} to expand $S^2$ and $S\ln S$ in~(\ref{eq:newODE}).   The expansion of $\ln S$ is obtained by first applying Cauchy's product rule to the identity $SS^{-1}=1$ and evaluating like-terms to obtain a recursive expression for the coefficients of the expansion of $S^{-1}$, given by~(\ref{eq:atilde}). The expansion of $S^{-1}$ is subsequently integrated term-by-term to obtain the expansion of $\ln S$, whose coefficients are given by~(\ref{eq:b}).  Although the series solution given by~(\ref{eq:SeriesSolution}) is an analytic solution to~(\ref{eq:newODE}), it is only valid within its radius of convergence and is incapable of capturing the long-time behavior of $S$.  This motivates the construction of an approximant to analytically continue the series beyond this convergence barrier.  


The long-time asymptotic behavior of the system~(\ref{eq:newODE}) is required to develop our asymptotic approximant, and so we proceed as follows.  It has been proven in prior literature~\cite{Hethcote} that $S$ approaches a limiting value, $S_\infty$, as $t\to\infty$, and this corresponds to $I\to0$ according to~(\ref{eq:newODE}). The value of $S_\infty$ satisfies equation~(\ref{eq:algebraic}) with $I=0$ as~\cite{Hethcote}
\begin{equation}
\frac{\alpha }{r} \ln \left(\frac{S_\infty }{S_0} \right)-S_{\infty } +S_{0} +I_{0} =0.  
\label{eq:infinite}
\end{equation} 

We expand $S$ as $t\to \infty $ as follows:
\begin{equation}
 S\sim S_{\infty } +S_{1} (t)\text{ where } S_{1} \to 0\text{ as }t\to \infty. 
\label{eq:perturb}
\end{equation}
Equation~(\ref{eq:perturb}) is substituted into~(\ref{eq:newODE}) and terms of O($S_{1} {}^{2} $) are neglected to achieve the following linearized equation
\begin{subequations}
 \begin{equation}
\frac{dS_{1} }{dt} =\kappa S_{1} 
\end{equation}
where
\begin{equation}
\kappa =rS_{\infty } -\alpha. 
\label{eq:kappa}
\end{equation}
\label{eq:asymptoticDE}
\end{subequations}
In writing~(\ref{eq:kappa}), the definition of $\beta$ in~(\ref{eq:beta}) has been employed.  Additionally, to obtain~(\ref{eq:asymptoticDE}), equation~(\ref{eq:infinite}) has been used which eliminates all O(1) terms in the linearized system.  The solution to~(\ref{eq:asymptoticDE}) is
\begin{equation}
 S_{1} =\varepsilon e^{\kappa t},
 \label{eq:S1}
 \end{equation}
where \textit{$\varepsilon$} is an unknown constant that can only be determined via connection with short-time behavior.  Consistent with the assumptions made, we find $\kappa <0$ such that $S_{1} \to 0$ as $t\to \infty $.  Thus the long-time asymptotic behavior of $S$ is given by
\begin{equation}
S\sim S_\infty+\varepsilon e^{\kappa t},~t\to\infty.
\label{eq:asymptotic}
\end{equation}
 Higher order corrections to the expansion~(\ref{eq:asymptotic}) may be obtained by the method of dominant balance~\cite{Bender} as a series of more rapidly damped exponentials of the form $e^{n\kappa t}$ where $n>1$.  This long-time asymptotic behavior of successive exponentials mimics that of the Sakiadis boundary layer problem describing flow along a moving plate in a stationary fluid~\cite{Barlow:2017}.  It is natural, then, to apply the Sakiadis approximant~\cite{Barlow:2017} to capture this asymptotic behavior while retaining the $t=0$ behavior given by~(\ref{eq:series}). The Sakiadis approximant imposes the exponential form of the long-time asymptotic behavior~(\ref{eq:asymptotic}) for all time; the coefficients of the exponentials are determined by matching their short-time expansion to the known power series developed about $t=0$ in the form of~(\ref{eq:series}).  However, here we find that a reciprocal expression that achieves the same $t\to\infty$ behavior~(\ref{eq:asymptotic}) (through its binomial expansion) converges faster than the original Sakiadis approixmant.

\begin{figure*}
\begin{tabular}{c}
\includegraphics[width=3.6in]{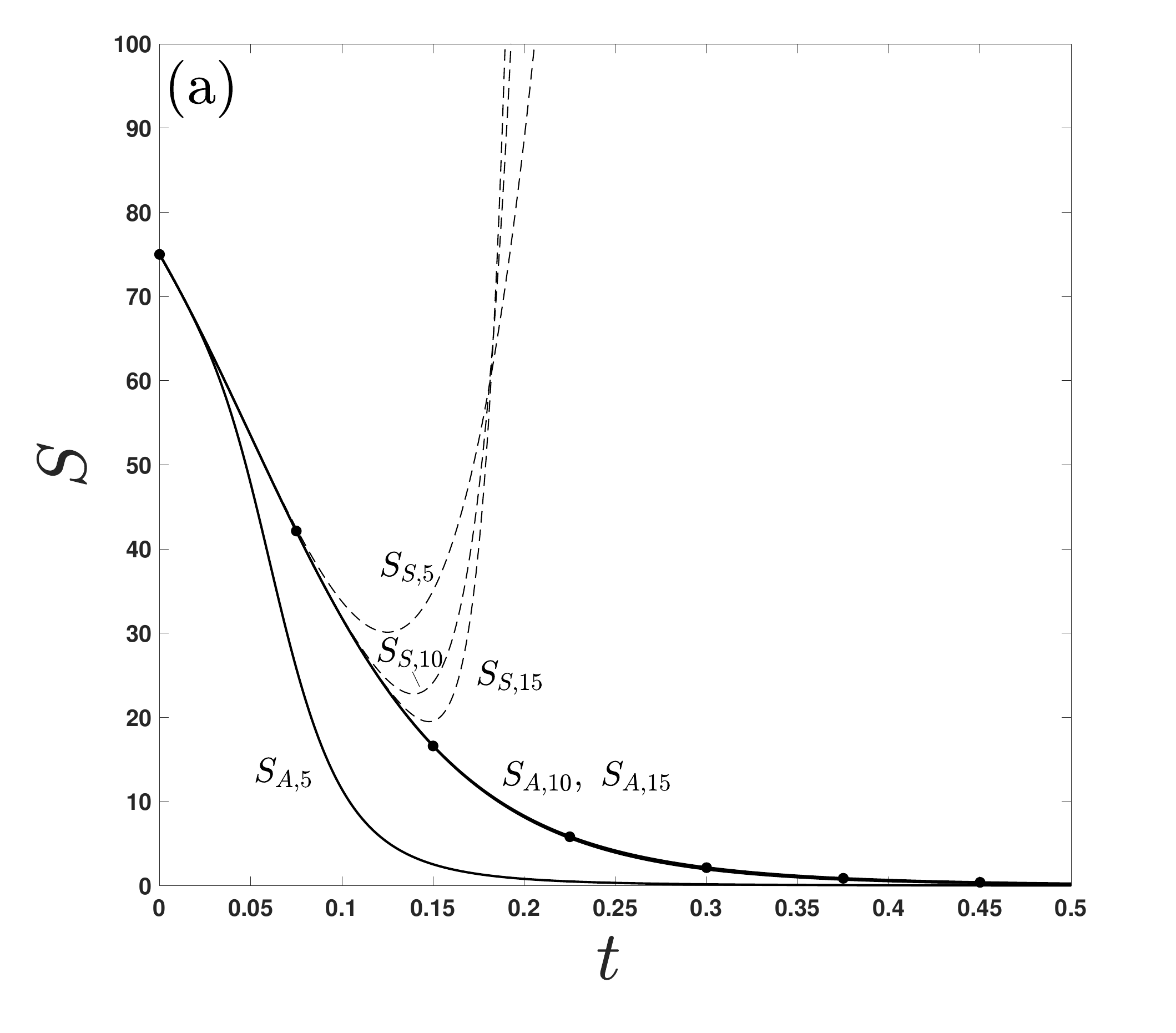} 
\includegraphics[width=3.6in]{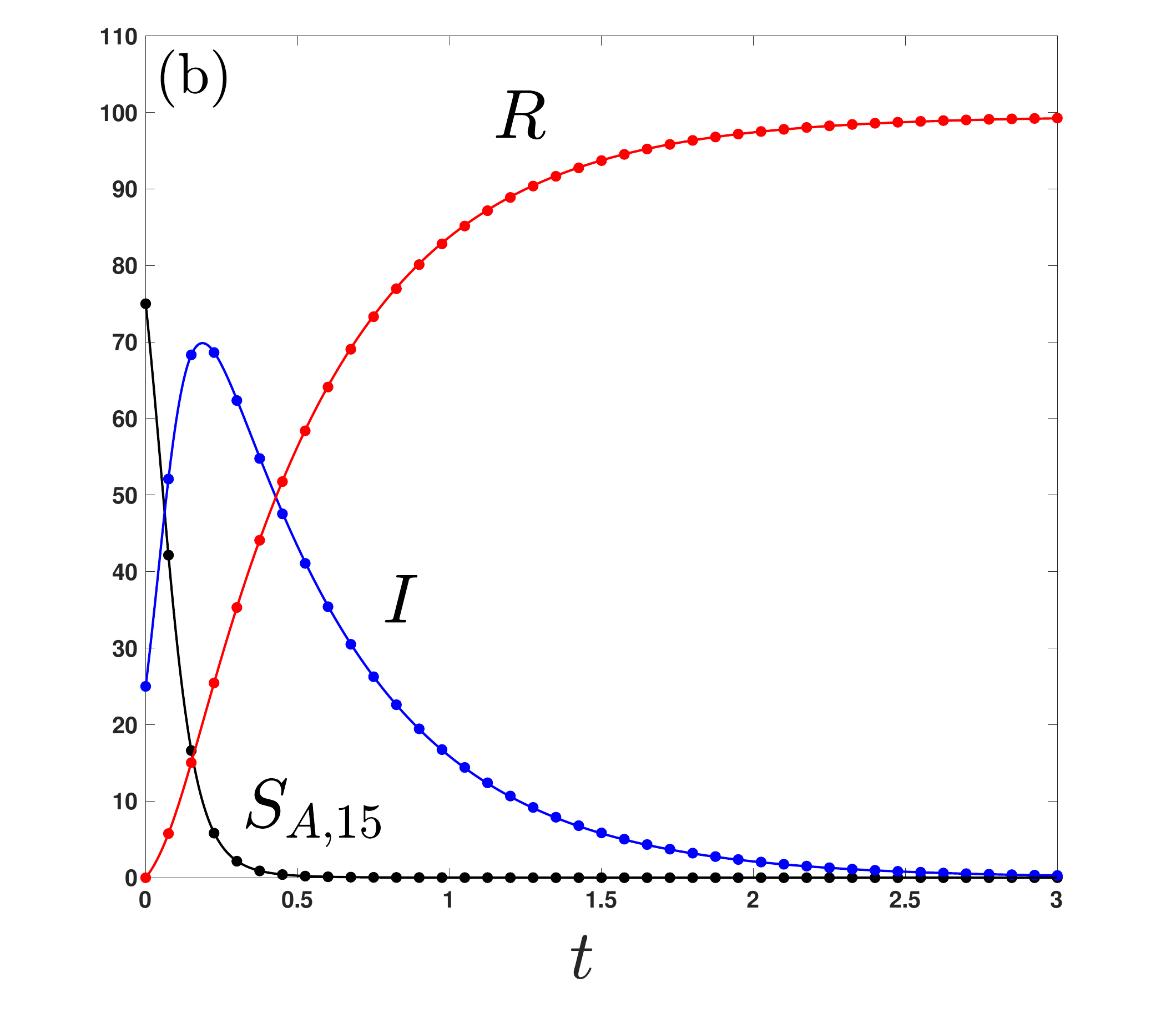} 
\end{tabular}
\caption{Analytical and numerical solutions to the SIR model~(\ref{eq:newODE}), where the susceptible ($S$), infected ($I$), and recovered ($R$) populations are plotted versus time, all in arbitrary units. (a) As the number of terms $N$ is increased, the series solution, denoted $S_{S,N}$ (given by~(\ref{eq:series}), dashed curves), diverges and the approximant, denoted $S_{A,N}$ (given by~(\ref{eq:approximant}), solid curves), converges to the exact (numerical) solution ($\bullet$'s).  (b) The converged asymptotic approximant for $S$ is used to obtain $R$ and $I$ (from equations~(\ref{eq:R}) and~(\ref{eq:algebraic}), respectively).  The model parameters values and initial conditions $\alpha=2$, $r=1/5$, $I_0=25$, and $S_0=75$ are taken from a test case used  in~\citet{Khan} to validate the homotopy analysis method.}
\label{fig:simple}
\end{figure*}

\begin{figure*}
\begin{tabular}{c}
\includegraphics[width=3.6in]{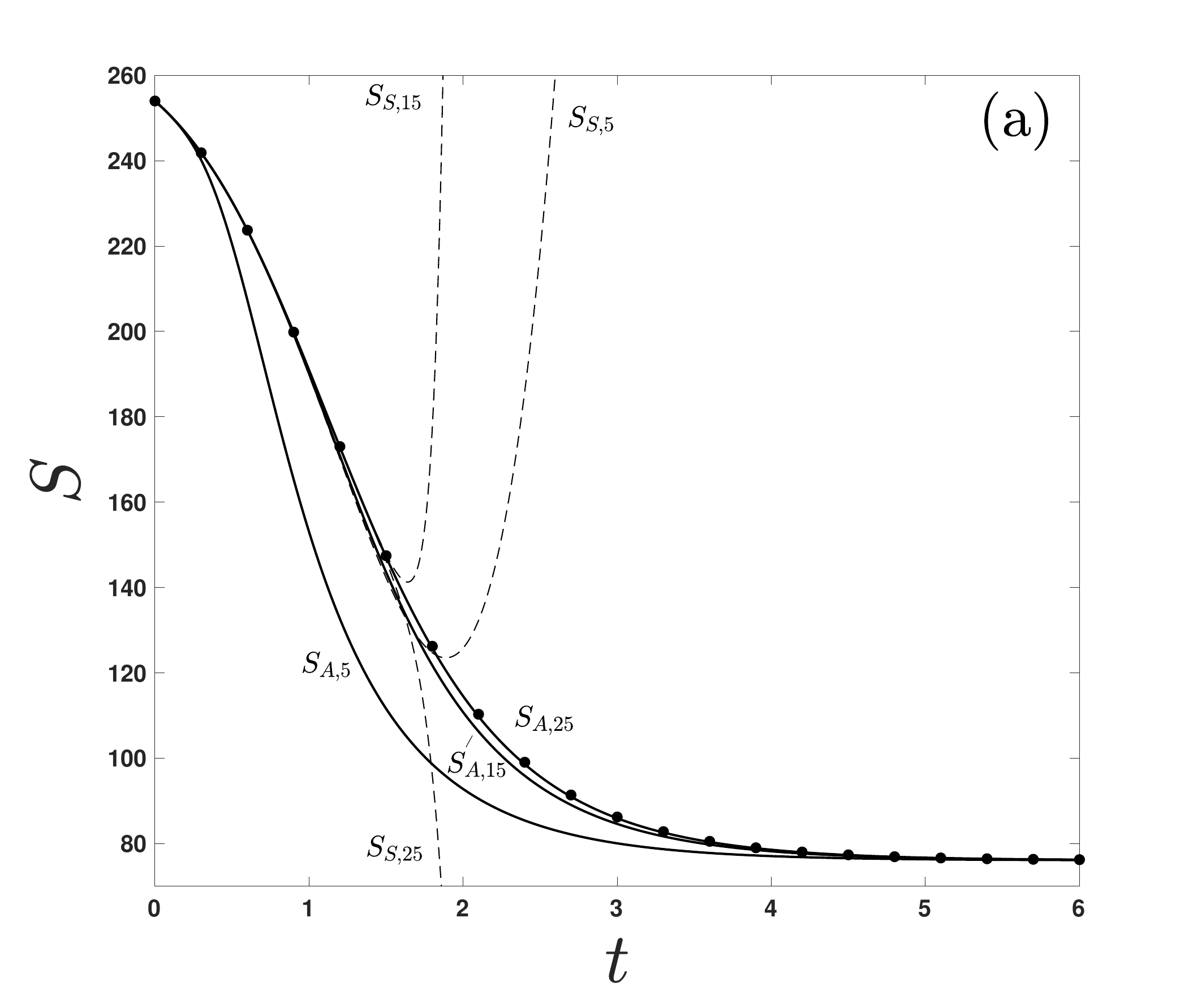} 
\includegraphics[width=3.6in]{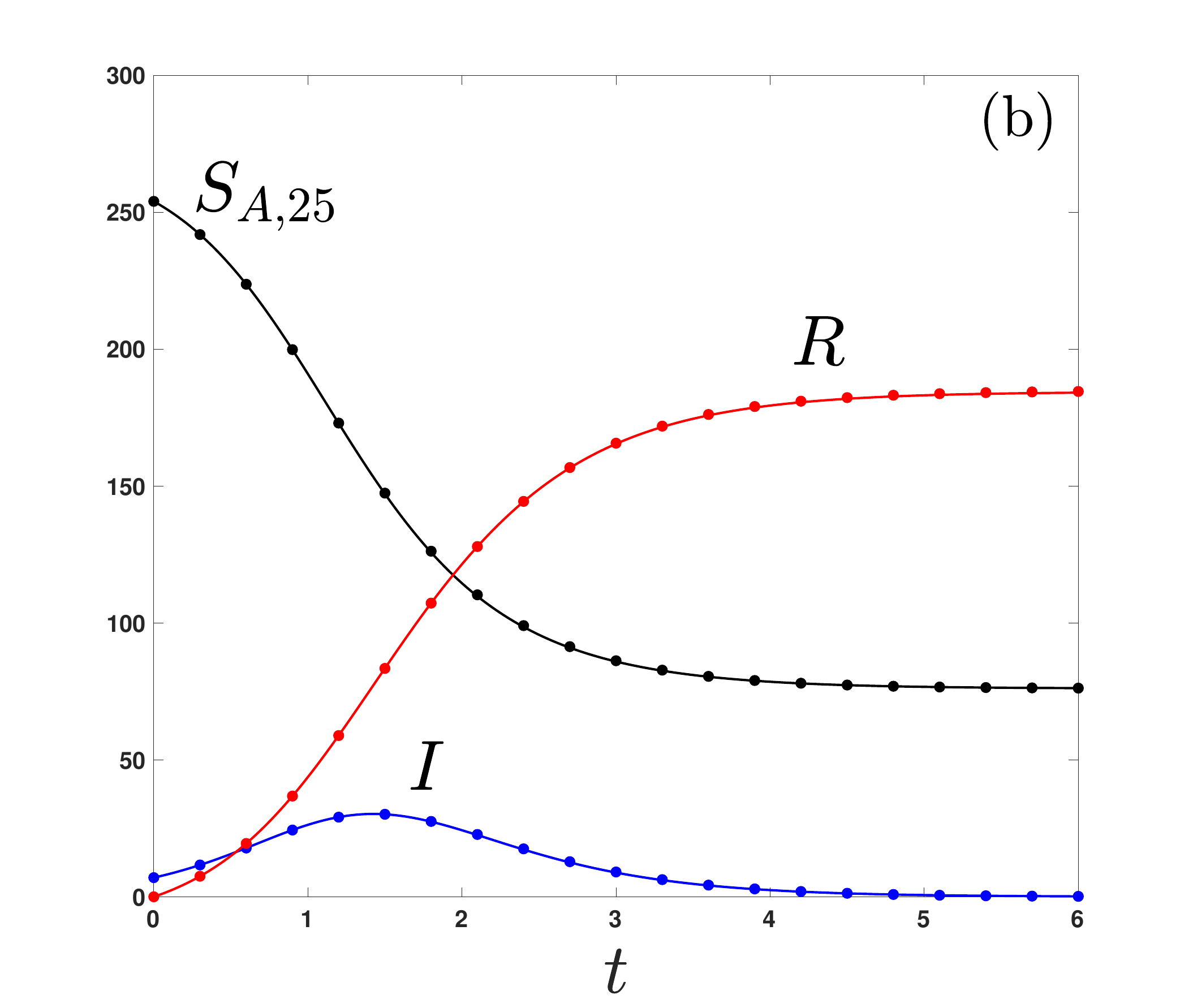} 
\end{tabular}
\caption{Analytical and numerical solutions to the SIR model~(\ref{eq:newODE}) where $S$, $I$, and $R$ are in units of people and $t$ is in months.  All other notation and labels are the same as in figure~\ref{fig:simple}. The model parameters values and initial conditions $\alpha=2.73$, $r=0.0178$, $I_0=7$, and $S_0=254$ are taken from estimates of the 1966 bubonic plague outbreak in Eyam, England examined in~\citet{Khan}}
\label{fig:bubonic}
\end{figure*}

\begin{figure*}
\begin{tabular}{c}
\includegraphics[width=3.6in]{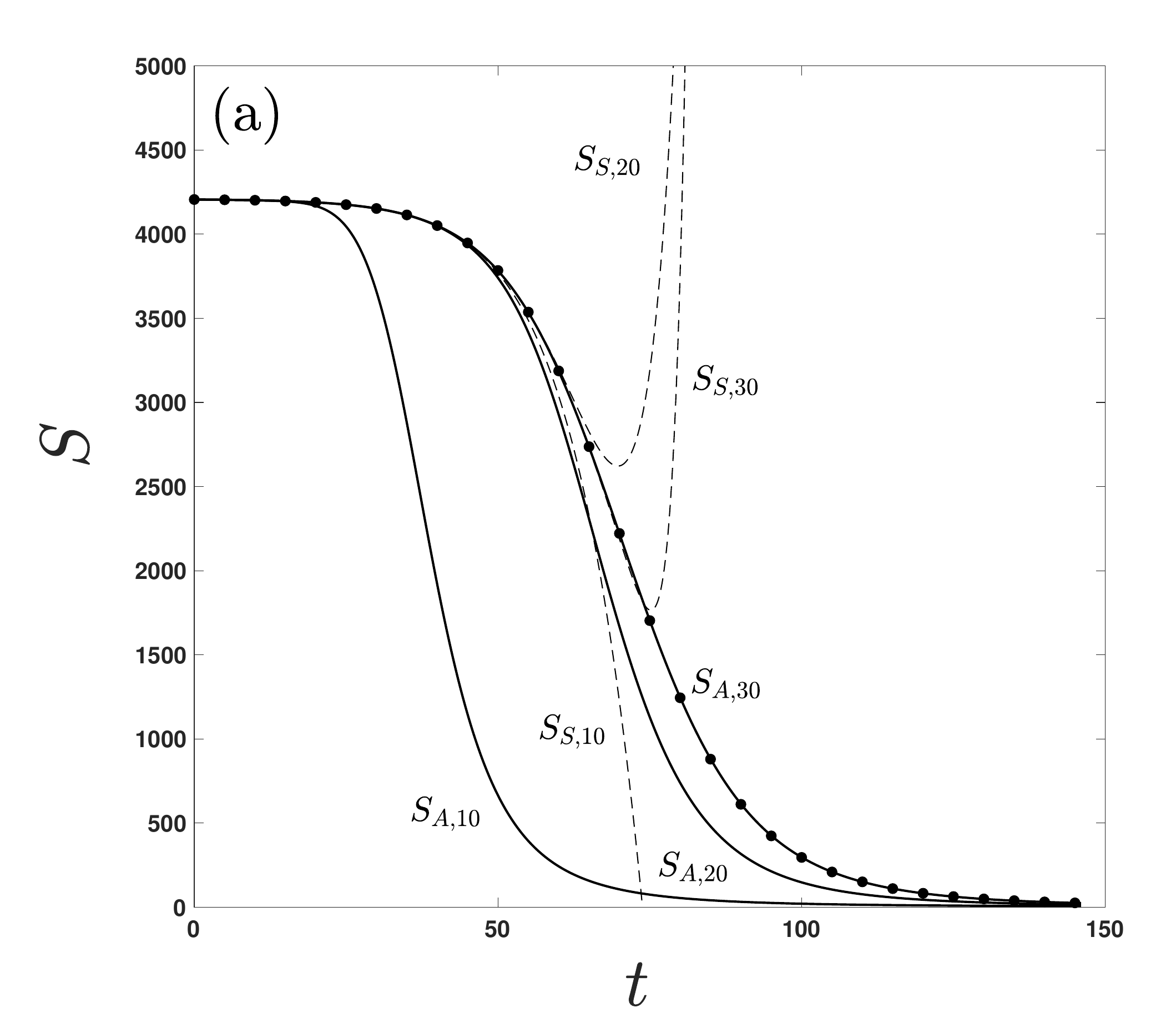} 
\includegraphics[width=3.6in]{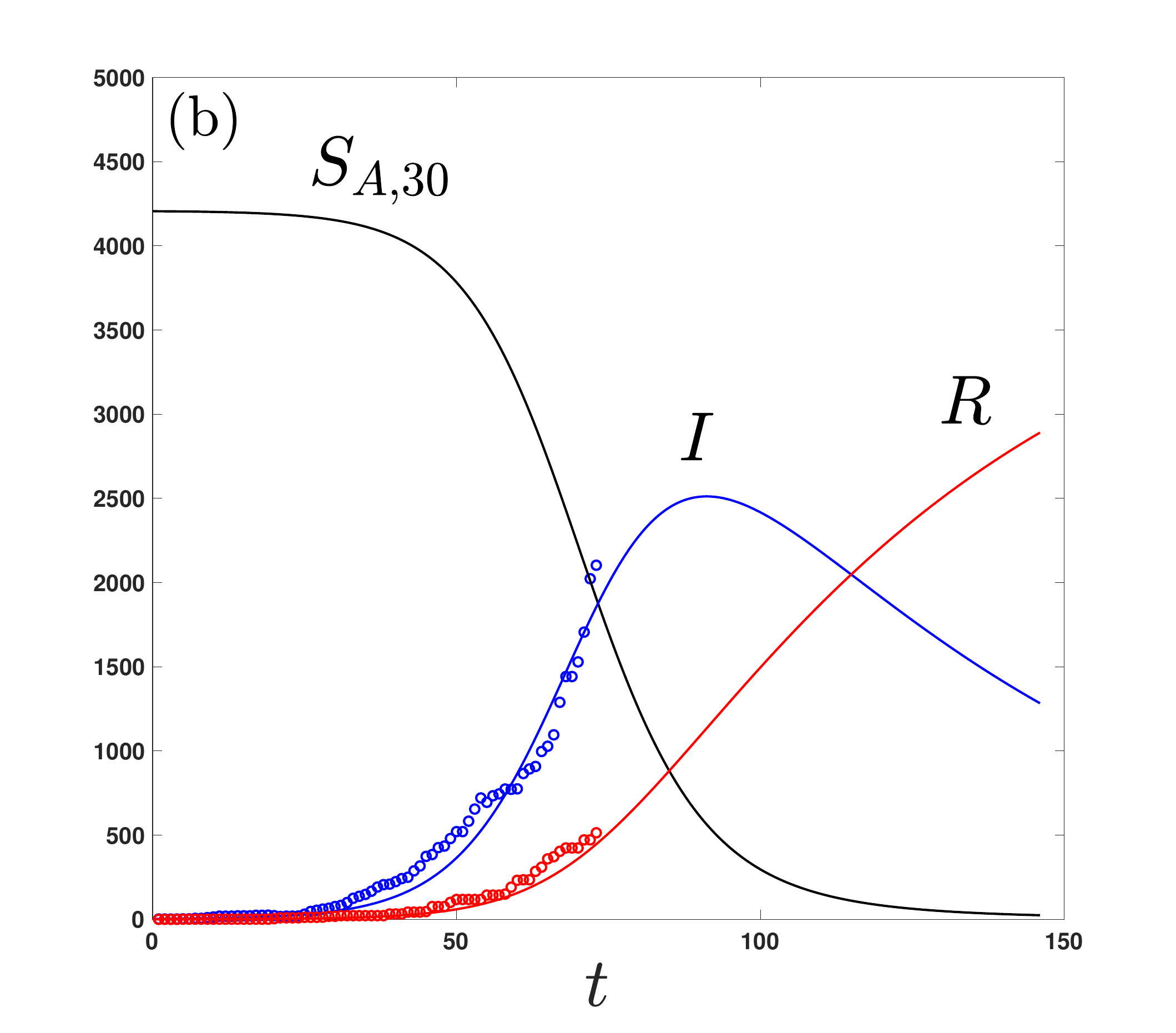} 
\end{tabular}
\caption{Analytical and numerical solutions to the SIR model~(\ref{eq:newODE}) where $S$, $I$, and $R$ are in units of people and $t$ is in days.  All other notation and labels are the same as in figure~\ref{fig:simple}. The model parameters values $\alpha$=0.0164 and $r$=2.9236$\times10^{-5}$ were obtained via a least-squares fit between the asymptotic approximant and  Japan COVID-19 outbreak data~\cite{covid} ($\circ$'s), using initial conditions  $I_0=2$ (from the first  point in the data set~\cite{covid}) and $S_0=4206$. Here $t=0$ is  January 22, 2020 (see main text for interpretation of the COVID-19 data).}
\label{fig:covid}
\end{figure*}

The assumed SIR approximant is given by 
\begin{subequations}
\begin{equation}
S_{A,N}=\frac{S_\infty}{1+\displaystyle\sum_{n=1}^N A_n e^{n \kappa t}} 
\label{eq:Aform}
\end{equation}
where the $A_n$'s are obtained by taking the reciprocal of both sides of~(\ref{eq:Aform}), expanding each side about $t=0$, and equating like-terms. The coefficients of the subsequent reciprocal expansion of the left-hand side (that of $S^{-1}$) are given by~(\ref{eq:atilde}).  After equating like-terms of this expansion with that of the reciprocal of the right-hand side of~(\ref{eq:Aform}), one arrives at the following linear system of equations to solve for the $A_n$ values as
\begin{equation}
\left[ \begin{array}{ccccc}
1^0 & 2^0 & 3^0 & \cdots & N^0 \\ 1^1 & 2^1 & 3^1 & \cdots & N^1 \\ 1^2 & 2^2 & 3^2 & \cdots & N^2 \\ \vdots & \vdots & \vdots & \vdots & \vdots \\ 1^{N-1} & 2^{N-1} & 3^{N-1} & \cdots & N^{N-1}\end{array} \right]\left[ \begin{array}{c}
A_1\\ A_2 \\ A_3 \\ \vdots \\ A_N\end{array} \right]=\vec{f},
\label{eq:matrix}
\end{equation}
\begin{equation}
\vec{f}=S_\infty\left[ \begin{array}{c}
0!~\tilde{a}_0-1/S_\infty\\ 1!~(1/\kappa)~\tilde{a}_1 \\ 2!~ (1/\kappa)^2 ~\tilde{a}_2 \\ \vdots \\ (N-1)!~ (1/\kappa)^{N-1} ~\tilde{a}_{N-1} \end{array} \right],
\label{eq:vector}
\end{equation}
\label{eq:approximant}
\end{subequations}
where~(\ref{eq:matrix}) is a Vandermonde matrix whose inversion is explicitly known~\cite{Turner:1966}.  The SIR approximant~(\ref{eq:approximant}) is thus a closed-form expression that, by construction, matches the correct $t\to\infty$ behavior given by~(\ref{eq:asymptotic}) and whose expansion about $t=0$ is exact to $N^\mathrm{th}$-order.  A MATLAB code for computing the $A_n$ coefficients is available from the authors~\cite{code}.  A Python code is also available from N. Barlow and J. Van Dyke\footnote[1]{https://github.com/nsbsma/SIR-approximant}.

Figure~\ref{fig:simple}a provides a typical comparison of the $N$-term series solution~(\ref{eq:SeriesSolution}) denoted by $S_{S,N}$ (and dashed lines), the $N$-term approximant~(\ref{eq:approximant}) denoted by $S_{A,N}$ (solid lines), and the numerical solution (indicated by symbols).  Note that the series solution has a finite radius of convergence as evidenced by the poor agreement and divergence from the numerical solution at larger times, even as additional terms are included.   By contrast, the approximant converges as additional terms are included.  For $N=15$, the approximant is visibly indistinguishable from the numerical solution (obtained by forward differencing) with a maximum relative error on the order of the numerical time-step (here $10^{-2}$) over the time range indicated.  Increasing the number of terms beyond $N=15$ does improve accuracy up to a point, but also increases the likelihood of deficient approximants for which the denominator can be zero for certain time values and specific values of $N$.  In general, the lowest number of terms that yields the desired accuracy is chosen to avoid this behavior.  The convergence of the approximant with increasing $N$ is a necessary condition for a valid approximant.  For the problems of mathematical physics to which we have applied asymptotic approximants~\cite{BarlowJCP,BarlowAIChE,Barlow2015,Barlow:2017,Barlow:2017b,Beachley,Belden}, we have observed that convergence of approximants implies excellent agreement with numerical results.  There is as-of-yet no proof developed that guarantees this result, but this interesting behavior has been a property of all approximants developed thus far.  In figure~\ref{fig:simple}b, the converged ($N=15$) asymptotic approximant for $S$ is used to obtain $R$ and $I$ (from equations~(\ref{eq:R}) and~(\ref{eq:algebraic}), respectively) and is compared with the numerical solution for these quantities. 

In figure~\ref{fig:bubonic}, the approximant is applied to a case examined in~\citet{Khan} to model the 1966 bubonic plague outbreak in Eyam, England. In figure~\ref{fig:covid}, the approximant is applied to COVID-19 data for Japan~\cite{covid}.  An increased number of terms in the approximant is required to achieve the same relative errors in figures~\ref{fig:simple},~\ref{fig:bubonic}, and~\ref{fig:covid}.  For all cases examined, we observe that this trend correlates with the breadth of the initial $S$ plateau.  

Note that the reported COVID-19 outbreak data~\cite{covid} in figure~\ref{fig:covid} is originally provided in terms of confirmed cases and recovered individuals per day.  The difference between these two quantities is used as an approximation to compare with the quantity $I$ of the SIR model.  It is acknowledged that the actual COVID-19 data is influenced by transient effects not included in the SIR model such as the exposure lag-time; these effects are incorporated in more sophisticated models such as SEIR~\cite{Hethcote}.  The approximation of $I$ from COVID-19 data is not restrictive in the current context, as our purpose is to show the efficacy of the closed form approximant rather than assess the validity of the SIR model.  

In figure~\ref{fig:covid}, a  least squares fit of the asymptotic approximant to $I$ data is used to extract SIR parameters $\alpha$ and $r$  based on data from the initial stages of the COVID-19 epidemic in Japan.  To do so,~(\ref{eq:algebraic}) is used to relate $I$ analytically to the solution for $S$ (here, the approximant $S_{A,30}$); note that $S_\infty$, used in the approximant, is affected by these parameters explicitly according to~(\ref{eq:infinite}).  The value of $S_0$ is not provided in the data set~\cite{covid}, and a least-squares algorithm is ineffective at determining an optimal value.  Here, we choose the value of $S_0$ to be twice that of the maximum value of $I$ approximated from the data, as it captures a typical curve shape for $S$ seen in applications of the SIR model~\cite{Hethcote}.  In regards to the  sensitivity of fitting parameters to the choice for $S_0$, a $100\%$ difference in $S_0$ leads to roughly a $50\%$ difference in $r$ and a $6\%$ difference in $\alpha$. The fit is made especially simple owing to the analytical form of the approximant that obviates the need to embed the numerical solution in such an algorithm.  The population of recovered individuals, $R$, is extracted from the solution for $I$ by direct integration in accordance with~(\ref{eq:R}).  Note that the predicted curve for $R$ in figure~\ref{fig:covid}, obtained solely by fitting data for $I$, is in good agreement with approximations from COVID-19 data for the recovered population, and serves as a check on the consistency of the data and algorithm.  

It is evident from the results presented here that an asymptotic approximant can be used to provide accurate analytic solutions to the SIR model.  Future work should focus on whether the asymptotic approximant technique can yield a closed form solution to more sophisticated epidemic models.



\end{document}